\documentclass[aps,twocolumn,prl,showpacs,superscriptaddress]{revtex4-1}
\usepackage{graphicx}
\usepackage{amsmath}
\usepackage{amsthm}
\usepackage{amssymb}
\usepackage{amsfonts}
\usepackage{siunitx}

\usepackage{blindtext}
\usepackage{comment}
\usepackage{xcolor}
\usepackage{xr}
\usepackage{lineno}
\usepackage[colorlinks]{hyperref}

\begin{document}
\title{Kovacs memory effect with an optically levitated nanoparticle}

\author{Andrei Militaru}\affiliation{Photonics Laboratory, ETH Z\"urich, CH-8093 Z\"urich, Switzerland}

\author{ Antonio Lasanta}\affiliation{Departamento de \'Algebra. Facultad 
de Educaci\'on, Econom\'ia y Tecnolog\'ia de Ceuta, Universidad de 
Granada, Cortadura del Valle, s/n, 51001 Ceuta, Spain}%
  \affiliation{Grupo de Teor\'{\i}as de Campos 
  y F\'{\i}sica Estad\'{\i}stica, Instituto Gregorio Mill\'an, 
  Universidad Carlos III de Madrid, Unidad Asociada al Instituto de 
  Estructura de la Materia, CSIC, Spain}%
  \affiliation{Grupo de Matem\'atica Aplicada a la F\'\i sica de la Materia Condensada, Instituto Gregorio Mill\'an, Universidad Carlos III de Madrid, Unidad Asociada al Instituto de Ciencias de Materiales de Madrid, CSIC, Spain}
\affiliation{Nanoparticles Trapping Laboratory, Universidad de Granada, 18071 Granada, Spain}  
  
  \author{Martin Frimmer}\affiliation{Photonics Laboratory, ETH Z\"urich, CH-8093 Z\"urich, Switzerland}

\author{Luis L. Bonilla}\affiliation{Departamento de Matem\'aticas, 
  Universidad Carlos III de Madrid, 28911 Legan\'es, Spain}%
\affiliation{Grupo de Matem\'atica Aplicada a la F\'\i sica de la Materia Condensada, Instituto Gregorio Mill\'an, Universidad Carlos III de Madrid, Unidad Asociada al Instituto de Ciencias de Materiales de Madrid, CSIC, Spain}  \affiliation{Instituto Gregorio Mill\'an, Universidad Carlos III de Madrid, 28911 Legan\'es, Spain}\

\author{Lukas Novotny}\affiliation{Photonics Laboratory, ETH Z\"urich, CH-8093 Z\"urich, Switzerland}

\author{Ra\'ul A. Rica}\affiliation{Nanoparticles Trapping Laboratory, Universidad de Granada, 18071 Granada, Spain}

\affiliation{Universidad de Granada, Department of Applied Physics and Research Unit “Modeling Nature” (MNat), 18071 Granada, Spain}

\begin{abstract}

The understanding of the dynamics of nonequilibrium cooling and heating processes at the nanoscale is still an open problem. These processes can follow surprising relaxation paths due to, e.g., memory effects, which significantly alter the expected equilibration routes. The Kovacs effect can take place when a thermalization process is suddenly interrupted by a change of the bath temperature, leading to a non-monotonic dependence of the energy of the system. Here, we demonstrate that the Kovacs effect can be observed in the thermalization of the center of mass motion of a levitated nanoparticle. The temperature is controlled during the experiment through an external source of white gaussian noise. We describe our experiments in terms of the dynamics of a Brownian particle in a harmonic trap without any fitting parameter, suggesting that the Kovacs effect can appear in a large variety of systems.  

\end{abstract}

\maketitle
\emph{Introduction.}---Memory effects (ME) are a unique feature of nonequilibrium systems \cite{KPZSN19}. Most of the ME appear as the transient evolution from a particular choice of initial conditions or as a response to a perturbation before the system relaxes to the equilibrium or stationary state. ME are ubiquitous in science, appearing in almost every field including physics, chemistry, biology, and materials science. They have been intensively studied in prototypical out of equilibrium systems such as disordered media \cite{BB02, janus:18, RKO03}, active matter \cite{JKL17}, and polymers \cite{struik80}, where ageing and rejuvenation are well understood. 

Thermalization processes are of utmost importance in many technological applications, and are prone to feature ME. As an example, the counterintuitive Mpemba effect has been recently rediscovered and attracted significant attention~\cite{MO69}. When the Mpemba effect takes place, it allows one to cool down faster the hotter of two systems (or heating up faster the cooler system). The Mpemba effect has been found in Markovian systems \cite{LR17,KRHM19}, granular matter \cite{LVPS17,Tal19}, spin glasses \cite{Bal19}, water \cite{GLH19},  the quantum Ising spin model \cite{NF19}, and very recently in a generalization to Markovian open quantum systems \cite{CLL21}. Under different conditions, the Kovacs effect (KE) \cite{K63,KAHR79} appears when trying to accelerate the cooling with a more elaborated protocol. Consider a system that at time $t=0$ is in equilibrium with a hot thermal reservoir at temperature $T_\mathrm{H}$ and that needs to be cooled down quickly to a warm temperature $T_\mathrm{W}<T_\mathrm{H}$. The Kovacs protocol first quenches the system by setting a cold temperature $T_\mathrm{C}<T_\mathrm{W}$, and then applies the target temperature $T_\mathrm{W}$ at a later time $t_\mathrm{W}$, when the system's temperature is already $T_\mathrm{W}$. Intuition tells us that with the Kovacs protocol the system reaches $T_\mathrm{W}$ faster than in a one step process. What actually happens, however, is that the system's temperature may not remain at $T_\mathrm{W}$: it may instead evolve following a non-monotonic path that features a hump.

KE and analogous crossover effects have been theoretically predicted in a large number of models, including spin glasses \cite{BB02}, glasses \cite{ALN06}, granular matter \cite{LVPS19,PT14}, active matter \cite{KSI17}, molecular gases \cite{MS04} and it has been generalized to athermal systems \cite{PP17}. As far as we know, however, the experimental observations are scarce. Apart from the original experiment with polymers \cite{K63,KAHR79}, the KE has been observed only in ferroelectrics \cite{PACD12} and granular matter \cite{JTMJ20}.

An optically trapped particle is an outstanding platform to experimentally explore nonequilibrium dynamics at the microscale~\cite{gieseler2014dynamic,martinez2016brownian}. Recent studies have engineered optimum protocols to exponentially speed up the thermalization process by taking advantage of the Mpemba effect \cite{kumar2020exponentially} and of the system's transient response upon a sudden change of a control parameter \cite{martinez2016engineered,chupeau2018engineered}. In this Letter, we demonstrate both experimentally and theoretically the occurrence of the KE in an optically levitated nanoparticle. We show that the the effect can be explained by the interplay of the different equilibration time scales predicted by the dynamics of a Brownian particle in a harmonic trap, anticipating that the same effect could be observed in a number of different situations.

\begin{figure}[htb]
	\centering
	\includegraphics[width=8.6cm]{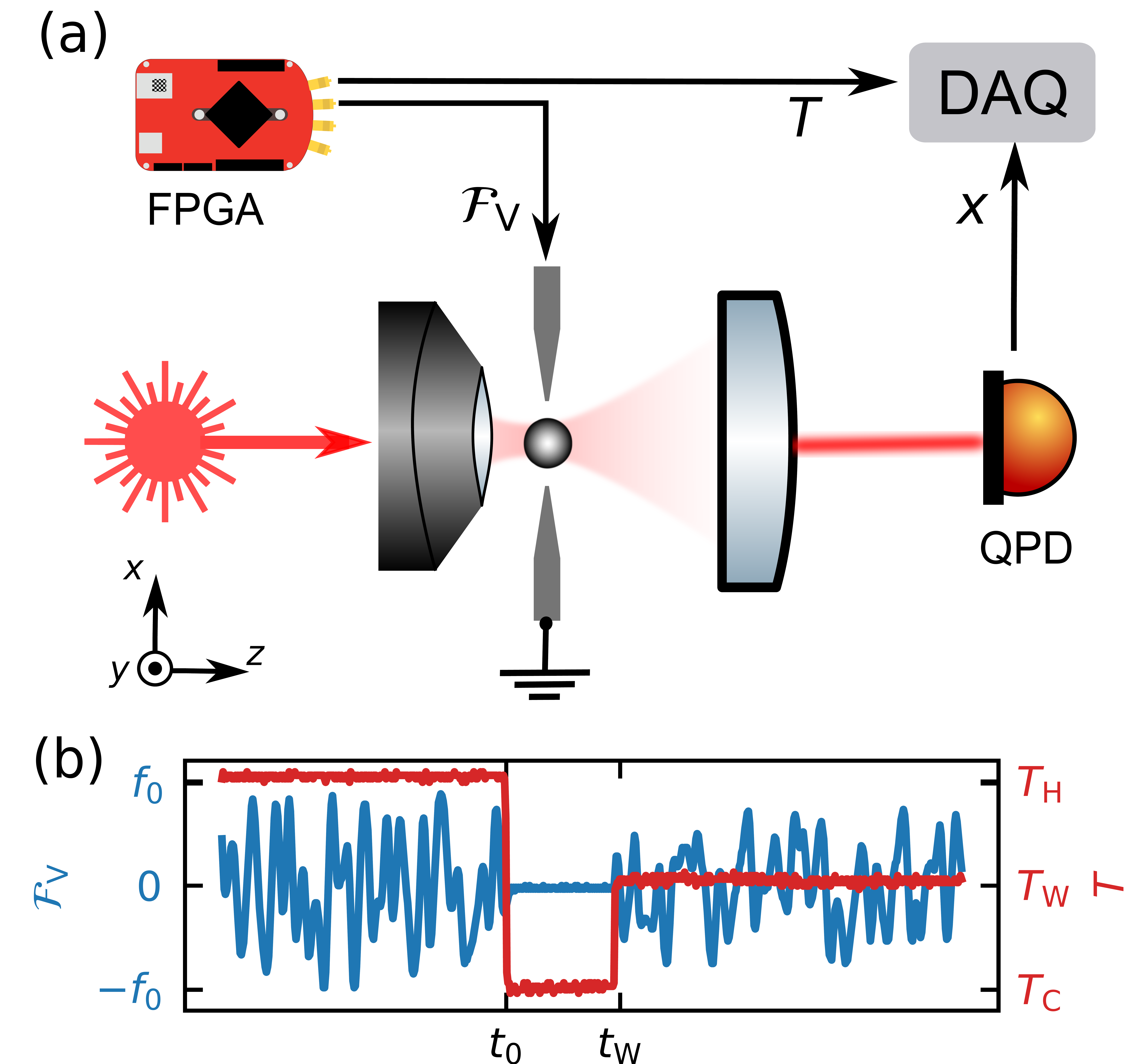}
	\caption{(a) A \SI{1064}{nm} wavelength laser is focused through a microscope objective (NA~0.8) to create an optical trap for a charged silica nanoparticle (radius \SI{90}{nm}). We use homodyne detection to record the particle's position with a quadrant photodiode (QPD). We use a custom programmed field programmable gate array (FPGA) to artificially tune the temperature of the particle along  the $x$ axis by applying a white gaussian electrostatic force. The instantaneous temperature $T$ is stored together with the position signal by a data acquisition card (DAQ). (b) Force $\mathcal{F}_\mathrm{v}$ applied on the particle by the FPGA and corresponding temperature $T$ during the Kovacs protocol shown in Fig.~\ref{fig:protocols}(b). $f_0$ is a reference force in arbitrary units. When $\mathcal{F}_\mathrm{v}=0$, the temperature coincides with the laboratory temperature $T_\mathrm{C} = \SI{298}{K}$. The values of the hot and of the warm temperatures are respectively $T_\mathrm{H} = \SI{2450}{K}$ and $T_\mathrm{W} = \SI{1290}{K}$.}
	\label{fig:setup}
\end{figure}

\emph{Experimental.}---Our experimental setup is shown in Fig.~\ref{fig:setup}(a). We create an optical potential for a  charged silica nanoparticle (radius $\approx \SI{90}{nm}$) by focusing an $x$-polarized laser beam traveling along the $z$ axis through a 0.8~NA microscope objective. We collect the trapping and the scattered light with a collection lens and apply a standard homodyne detection scheme to measure the position of the particle \cite{Gieseler2012, Hebestreit2018a}. We use a pair of electrodes mounted along the $x$ axis to control the motion of the particle through an externally applied electrostatic force \cite{martinez2013effective,Frimmer2017d}. To first order, the particle obeys the following equation of motion:
\begin{equation}
\label{eq: EOM}
m\ddot{x} + m\Gamma_0 \dot{x} + m \Omega_0^2 x = \mathcal{F}_\mathrm{th} + \mathcal{F}_\mathrm{v},
\end{equation}
with analogous expressions for the other two axes $y$ and $z$. In Eq.~\eqref{eq: EOM},  $m\approx\SI{6.6}{fg}$ is the mass of the particle, $\Omega_0=2\pi\times \SI{159(1)}{kHz}$ is the resonance frequency of the optical trap and $\Gamma_0 = 2\pi\times \SI{399(2)}{kHz}$ is the damping rate due to collisions with gas molecules. By virtue of the fluctuation-dissipation theorem, the damping $\Gamma_0$ leads to a fluctuating force $\mathcal{F}_\mathrm{th}$ with $\langle\mathcal{F}_\mathrm{th}(t) \rangle=0$ and $\langle\mathcal{F}_\mathrm{th}(t)\mathcal{F}_\mathrm{th}(t+\tau) \rangle= 2m k_\mathrm{B}T_\mathrm{C} \Gamma_0~\delta(\tau)$, where $T_\mathrm{C} = \SI{298}{K}$ is the laboratory temperature and $\langle \cdot \rangle$ refers to the ensemble average \cite{Kubo1966}. Note that our system lives in the largely unexplored damping regime that is neither underdamped ($\Gamma_0 < 2 \Omega_0$) nor deeply overdamped ($\Gamma_0 \gg \Omega_0$). By adding an external stochastic force $\mathcal{F}_\mathrm{v}$ we increase the strength of the fluctuation without affecting the dissipation, such that the net effect is to tune the effective temperature of the center of mass motion (COM) \cite{martinez2013effective}. The COM temperature is controlled in time by a custom programmed field programmable gate array (FPGA) that tunes the variance of $\mathcal{F}_\mathrm{v}$, as shown in Fig.~\ref{fig:setup}(b). The bandwidth of $\mathcal{F}_\mathrm{v}$ is \SI{5}{MHz}, such that it can be considered a white gaussian force for all purposes. In the rest of this Letter, we make use of three values of the COM temperature: (\emph{i}) cold (or laboratory) temperature $T_\mathrm{C}$, (\emph{ii}) warm temperature $T_\mathrm{W}=\SI{1290}{K}$, and (\emph{iii}) hot temperature $T_\mathrm{H}=\SI{2450}{K}$.

\begin{figure}[htb]
	\centering
	\includegraphics[width=8.6cm]{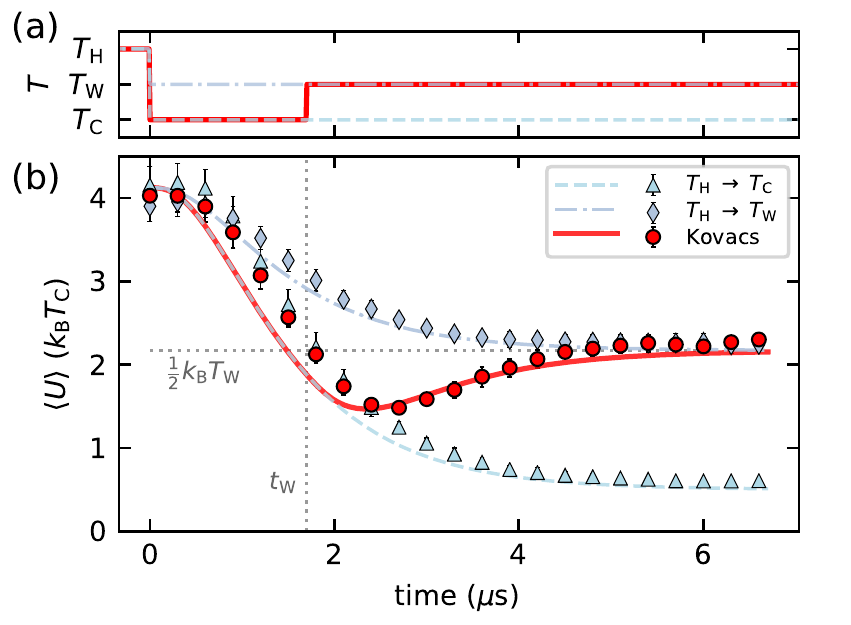}
	\caption{(a) Temperature of the center of mass motion (COM) during the cooling protocol $T_\mathrm{H} \rightarrow T_\mathrm{C}$ (dashed light blue line), cooling protocol $T_\mathrm{H} \rightarrow T_\mathrm{W}$ (dash-dotted silver line) and Kovacs protocol (red solid line). (b) Average potential energy during the protocols described in (a). Discrete symbols (light blue triangles, silver diamonds and red circles) are used for experimental data while lines are used for the theoretical curves. No free parameters have been used in the model. The horizontal dotted line is the equilibrium value $k_\mathrm{B}T_\mathrm{W}/2$ related to the warm temperature. The vertical dotted line signals the switch time between $T_\mathrm{C}$ and $T_\mathrm{W}$ for the Kovacs protocol data. The estimation of the error bars is described in the Supplementary Material.}
	\label{fig:protocols}
\end{figure}

\emph{Results.}---The thermodynamic quantity of interest is the particle's potential energy, defined as $\langle U \rangle = m\Omega_0^2 \langle x^2 \rangle/2$. We measure the time evolution of $\langle U \rangle$ by computing the ensemble average over $10000$ trajectories of the particle that start from the same equilibrium state. Figure~\ref{fig:protocols} displays the evolution of the potential energy and the corresponding changes in temperature for three different protocols.

The first protocol is the equilibration from the hot to the cold temperature (light blue triangles). This quenching of the thermal bath from $T_\mathrm{H}$ to $T_\mathrm{C}$ induces an equilibration of the potential energy of the particle that takes about $\SI{4}{\mu s}$. Interestingly, rather than decaying exponentially towards $k_\mathrm{B}T_\mathrm{C}/2$, the potential energy remains at $k_\mathrm{B}T_\mathrm{H}/2$ for a duration of $1/\Gamma_0 \approx \SI{400}{ns}$ before decaying. In general, the potential energy of an underdamped (overdamped) oscillator would follow a simple exponential decay with rate given by $\Gamma_0$ ($\Omega_\mathrm{c}=\Omega_0^2/\Gamma_0$) \cite{Tebbenjohanns2019,kumar2020exponentially}. The nonexponential equilibration of $\langle U \rangle$ for our intermediate damping regime suggests that more than one time scale is involved in the particle's equilibration. 

The second protocol we analyze is the equilibration between hot and warm temperature (silver diamonds). The $T_\mathrm{H} \rightarrow T_\mathrm{W}$ equilibration exhibits the same qualitative behavior of $T_\mathrm{H} \rightarrow T_\mathrm{C}$ except for the final state, which in this case is $k_\mathrm{B}T_\mathrm{W}/2$. During the $T_\mathrm{H} \rightarrow T_\mathrm{C}$ equilibration, the intermediate value of $k_\mathrm{B}T_\mathrm{W}/2$ is reached after a time $t_\mathrm{W} = \SI{1.7}{\mu s}$.

Finally, we apply the Kovacs protocol (red data points in Fig~\ref{fig:protocols}): (\emph{i}) we let the system equilibrate at $T_\mathrm{H}$, (\emph{ii}) at time $t=0$ we change the temperature to $T_\mathrm{C}$, (\emph{iii}) at time $t_\mathrm{W}$ we switch the temperature to $T_\mathrm{W}$ \cite{PT14, Bertin2003, Prados2010}. Despite the fact that $\langle U \rangle$ is already at its correct steady state value at $t=t_\mathrm{W}$, its value keeps following the $T_\mathrm{H} \rightarrow T_\mathrm{C}$ equilibration curve and only later reverts back towards $k_\mathrm{B}T_\mathrm{W}/2$. The observed behavior is known as the \emph{anomalous Kovacs effect} \cite{PT14}, and this is to the best of our knowledge its first experimental observation with a single particle. In the Supplementary Material, we show that the very same behaviour is observed when the initial state is prepared at $T_\mathrm{C}$ and a heating Kovacs protocol is applied to reach $T_\mathrm{W}$.

We resort to an analytical solution of Eq.~\eqref{eq: EOM} to explain the measurements. As detailed in the Supplementary Material \cite{supplemental}, the equilibration of the potential energy can be described by the following equation:
\begin{subequations}
    \label{eq: equilibration}
    \begin{align}
        \langle \tilde{U} \rangle &= 1 + a_1 e^{-2\lambda_1 t} + a_2 e^{-2\lambda_2 t} + a_3 e^{-\Gamma_0 t}, \label{eq: exponentials}\\
        \lambda_1 &= \frac{1}{2}\left(\Gamma_0 + \sqrt{\Gamma_0^2 - 4\Omega_0^2}\right), \\
        \lambda_2 &= \frac{1}{2}\left(\Gamma_0 - \sqrt{\Gamma_0^2 - 4\Omega_0^2}\right),
    \end{align}
\end{subequations}
where $\tilde{U}$ is the potential energy normalized to its final value $k_\mathrm{B}T_\mathrm{f}/2$ ($T_\mathrm{f}$ being the final temperature), $\lambda_1$ and $\lambda_2$ are the two poles of Eq.~\eqref{eq: EOM}, and $\{a\}=\{a_1, a_2, a_3\}$ are coefficients that depend on the initial and final states and whose explicit expressions are detailed in the Supplementary Material. The equilibrations $T_\mathrm{H} \rightarrow T_\mathrm{C}$ and $T_\mathrm{H} \rightarrow T_\mathrm{W}$ can be computed in a straightforwards fashion by substituting the values of the $\{a\}$ coefficients that correspond to the specific initial and final thermal states. For the case of the Kovacs protocol, we must compute the equilibration in three steps: (\emph{i}) we use the equilibration $T_\mathrm{H} \rightarrow T_\mathrm{C}$ up to $t=t_\mathrm{W}$, (\emph{ii}) we use the instantaneous value $\langle U \rangle (t_\mathrm{W})$ and the final temperature $T_\mathrm{W}$ to determine the new values of the $a$ coefficients and (\emph{iii}) compute the rest of the equilibration. The dashed, dash-dotted and continuous lines in Fig.~\ref{fig:protocols}(b) represent the theoretical equilibrations computed from Eq.~\eqref{eq: equilibration}, which are in good agreement with the experiment. All quantities have been estimated independently and no free parameter has been used.

\begin{figure*}[htb]
	\centering
	\includegraphics[width=17.8cm]{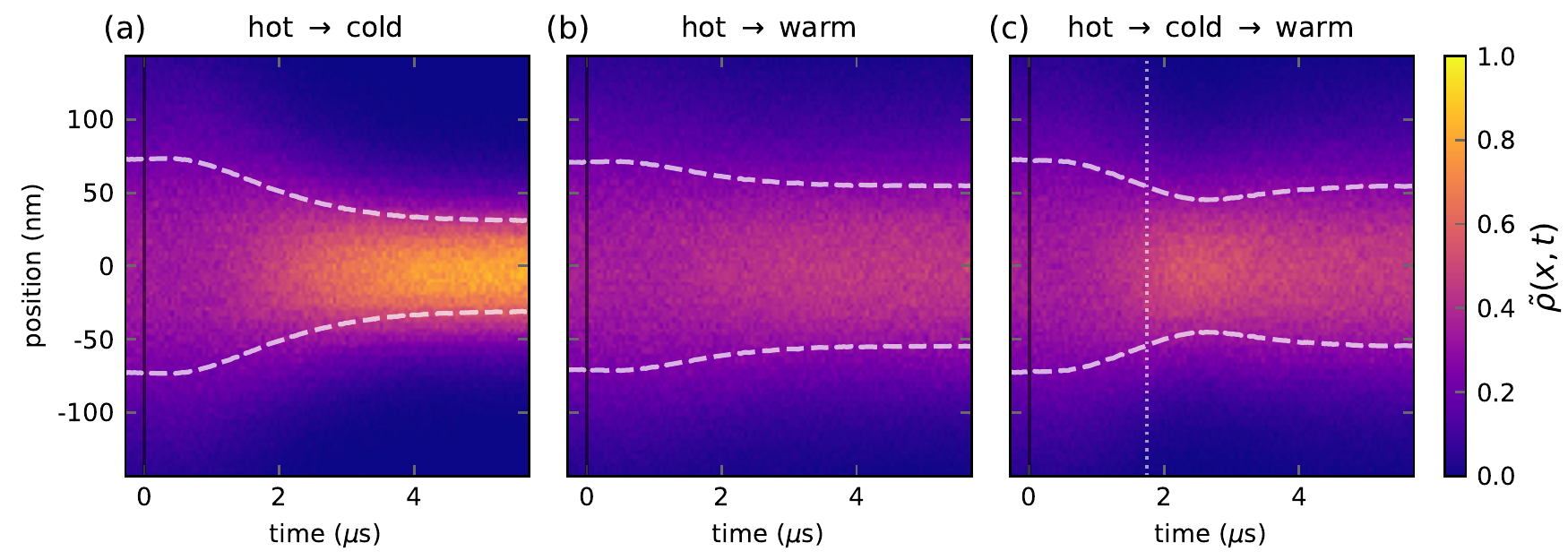}
	\caption{(a) Measured position distribution $\tilde{\rho}(x, t)$ of the particle during the equilibration $T_\mathrm{H} \rightarrow T_\mathrm{C}$, (b) $T_\mathrm{H} \rightarrow T_\mathrm{W}$, (c) $T_\mathrm{H} \rightarrow T_\mathrm{C} \rightarrow T_\mathrm{W}$ (Kovacs protocol). The vertical black line indicates the time $t=0$ where the temperature is first changed. In (c), the vertical dotted white line shows the switch time $t=t_\mathrm{W}$ between $T_\mathrm{C}$ and $T_\mathrm{W}$. The dashed white lines represent the standard deviations inferred from Fig.~\ref{fig:protocols}. At all times, the distributions are gaussian. All distributions are normalised to the maximum value of (a).}
	\label{fig:histograms}
\end{figure*}

In order to gain deeper insight into the system's equilibration, we use the measured trajectories to estimate the position's probability distribution $\tilde{\rho}(x,t)$ as a function of time. The distributions for each of the three equilibration protocols are shown in Fig.~\ref{fig:histograms}. The position distributions exhibit a gaussian shape at all instants of time. Specifically, during the Kovacs protocol, the distribution at the switch time $t_\mathrm{W}$ is the same as in the steady state $t \to \infty$. The fact that the distribution keeps narrowing for $t > t_\mathrm{W}$, however, implies that the particle's state is actually a nonequilibrium one. The solution to this apparent contradiction lies in the fact that Eq.~\eqref{eq: EOM} has actually two state variables, position $x$ and velocity $v$. The measurements in Fig.~\ref{fig:histograms} describe the projection of the total probability distribution $\rho(x,v)$ over the $x$ axis, not the full state of the particle. 

\begin{figure}[htb]
	\centering
	\includegraphics[width=8.6cm]{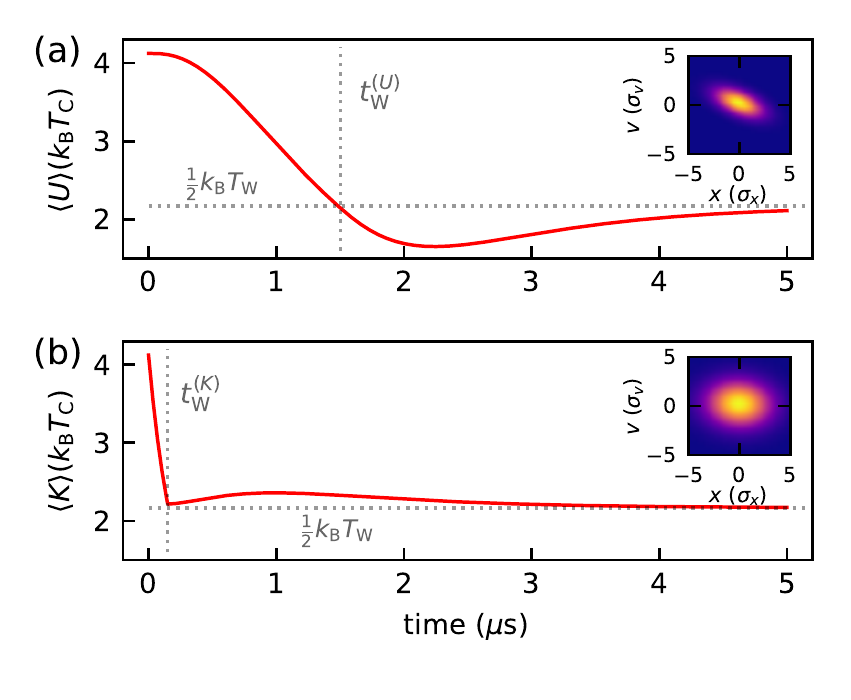}
	\caption{(a) Theoretical Kovacs equilibration of the potential energy. At $t=0$ the temperature is switched from $T_\mathrm{H}$ to $T_\mathrm{C}$. At time $t_\mathrm{W}^{(U)}$ (grey dotted line), the temperature is switched to $T_\mathrm{W}$. Inset: phase space distribution at $t=t_\mathrm{W}^{(U)}$. The distribution is classically squeezed and has a nonzero  $\langle xv \rangle$ correlation. (b) Theoretical Kovacs effect of the kinetic energy. The switch time (grey dotted line) occurs at $t = t_\mathrm{W}^{(K)}$ which is general different from $t_\mathrm{W}^{(U)}$. Inset: phase space distribution at $t=t_\mathrm{W}^{(K)}$, where a different nonequilibrium state that the one in (a) is visible. Both insets are normalized to their respective maximum and the same colorbar as in Fig.~\ref{fig:histograms} is used.}
	\label{fig:theory}
\end{figure}

In order to reconstruct the evolution of $\rho(x, v, t)$, analogue formulas to Eq.~\eqref{eq: equilibration} can be derived for the kinetic energy and for the correlation $\langle xv \rangle$. Figure~\ref{fig:theory} shows the theoretical curves of the Kovacs effect applied both to the potential and to the kinetic energies. Since the two energies follow different time evolutions, the Kovacs protocol cannot be applied to both quantities simultaneously, so we use two separate switch times $t_\mathrm{W}^{(U)}$ and $t_\mathrm{W}^{(K)}$. Remarkably, when the protocol is applied to the kinetic energy, a standard Kovacs effect is observed. It appears thus that a harmonic oscillator at intermediate dampings can exhibit either anomalous or standard Kovacs effect depending on the thermodynamic quantity under study. 

 The two insets in Fig.~\ref{fig:theory} represent two snapshots of $\rho$ taken at the Kovacs switch times $t_\mathrm{W}^{(U)}$ and $t_\mathrm{W}^{(K)}$. The distribution is gaussian and presents classical squeezing. In the case of $\langle U \rangle$, in addition to the classical squeezing there is a visible anticorrelation $\langle xv \rangle$. Both intermediate distributions are nonequilibrium states, which explains the equilibrations in Fig.~\ref{fig:protocols} despite the gaussian distributions of Fig.~\ref{fig:histograms}.
 
 \emph{Discussion.}---In an overdamped harmonic oscillator, one typically considers the dynamics of the particle as governed by a single time scale given by the cutoff frequency $\Omega_\mathrm{c} = \Omega_0^2/\Gamma_0$ \cite{martinez2016engineered,kumar2020exponentially}. However, a single time scale governing the equilibration could not generate nonequilibrium states like the ones in Fig.~\ref{fig:theory}, but rather only states that correspond to some intermediate temperature. A second, intrinsic time scale of the system is given by the damping coefficient $\Gamma_0$. The particle exhibits ballistic motion when its evolution is resolved within intervals on the order of $1/\Gamma_0$, such that the overdamped approximation is never fully valid \cite{Li2010}. In the intermediate regime of our experiment, the ballistic and the cutoff time scales are close to each other. The result is that the equilibration of the potential energy is given by a superposition of three exponentials with time constants derived from both $\Gamma_0$ and $\Omega_\mathrm{c}$. The superposition of these exponentials is the key to explain the measured Kovacs effect. In the case of the potential energy, the particle cannot respond instantaneously to the changes in temperature because of the ballistic motion over short time intervals. This inertia-like effect forces $\langle U \rangle$ during the Kovacs protocol to continue its decreasing trend even after the switch to $T_\mathrm{W}$, generating thus an anomalous hump. Unlike $\langle U \rangle$, the changes of the kinetic energy are instantly affected by the the changes in temperature, which is the reason why Fig.~\ref{fig:theory}(b) presents kink points. At the switching time $t_\mathrm{W}^{(K)}$, the kinetic energy has the correct steady state value, but the potential energy is higher. In order to reach equilibrium with $T_\mathrm{W}$, the potential energy must first be converted to kinetic energy and then dissipated. It is this transfer from potential to kinetic energy that gives rise to the standard Kovacs hump of Fig.~\ref{fig:theory}(b). 
 
 In contrast to the overdamped case, when the particle experiences a low friction $\Gamma_0 \ll \Omega_0$ one finds a new separation of the effects of the two different time scales. On the one hand, the resonance frequency $\Omega_0$ induces coherent oscillations in the motion of the particle. On the other hand, the friction $\Gamma_0$ induces random fluctuations and dissipates the excess energy. For this deeply underdamped regime, the first two exponentials of Eq.~\eqref{eq: exponentials} become vanishingly small and we recover once again an equilibration given by one time scale only, $\Gamma_0$ \cite{supplemental}. The Kovacs effect in the harmonic oscillator is thus a result of the intermediate damping regime, which has been so far only marginally explored. 

\emph{Conclusions.}---We have demonstrated that the energy equilibration of the harmonic oscillator can exhibit the Kovacs effect. This observation on one of the most widely used models in physics, the harmonic oscillator, highlights the generality and the importance of this phenomenon across fields. A deep understanding of the dynamics of thermalization processes performed out of equilibrium is required in many situations, and the exact mathematical solution developed in this work provides the opportunity to recognise what features of the Kovacs effect present in literature are general or resulting from an approximation.  We emphasize that our experiments have been performed in the overdamped \emph{regime} ($\Gamma_0^2>4\Omega_0^2$), but under a situation where the frequently used overdamped \emph{approximation}, i.e., neglecting the first addend on the \emph{lhs} term of Eq.~\ref{eq: EOM}, cannot be applied. This warns against the adoption of such an approximation under certain circumstances, because it may return erroneous conclusions and predictions, as has been previously observed in similar situations of low damping~\cite{pan2018quantifying} or when the temperature of the system is allowed to evolve with time~\cite{roldan2014measuring,martinez2015adiabatic}. The accurate quantification of time scales in thermalization processes of Brownian systems is also required for the correct evaluation of performances and efficiencies in the implementation of thermodynamics processes and thermal machines at the microscale, the so called stochastic thermodynamics~\cite{seifert2012stochastic,martinez2017colloidal,gieseler2018levitated}.

\emph{Acknowledgements.}--- This research has been supported by European Union’s Horizon 2020 research and innovation programme under Grant No. 863132 (iQLev) and by the Swiss National Science Foundation through Grant No. 200021L-169319. AL and LLB acknowledge financial support by the FEDER / Ministerio de Ciencia, Innovaci\'on y Universidades -- Agencia Estatal de Investigaci\'on, under grant MTM2017-84446-C2-2-R, LLB acknowledges financial support by the Madrid Government (Comunidad de Madrid-Spain) under the Multiannual Agreement with UC3M in the line of Excellence of University Professors (EPUC3M23), and in the context of the V PRICIT (Regional Programme of Research and Technological Innovation). RAR acknowledges financial support from FEDER/Junta de Andalucía-Consejería de Economía y Conocimiento/Projects C-FQM-410-UGR18 and P18-FR-3583.

\bibliographystyle{apsrev4-1}

\end{document}